\documentclass[twocolumn,showpacs,prl]{revtex4}

\ifx\pdfoutput\undefined
\usepackage{graphicx}
\else
\usepackage[pdftex]{graphicx}
\usepackage{epstopdf}
\fi

\usepackage[center]{subfigure}

\usepackage{color}

\begin{document}

\title{Microscopic dissipation in a cohesionless granular jet impact}

\author{Nicholas Guttenberg}
\affiliation{James Franck Institute}

\begin{abstract}


Sufficiently fine granular systems appear to exhibit continuum properties, though the precise continuum limit obtained can be vastly different depending on the particular system. We investigate the continuum limit of an unconfined, dense granular flow. To do this we use as a test system a two-dimensional dense cohesionless granular jet impinging upon a target. We simulate this via a timestep driven hard sphere method, and apply a mean-field theoretical approach to connect the macroscopic flow with the microscopic material parameters of the grains. We observe that the flow separates into a cone with an interior cone angle determined by the conservation of momentum and the dissipation of energy. From the cone angle we extract a dimensionless quantity $A-B$ that characterizes the flow. We find that this quantity depends both on whether or not a deadzone --- a stationary region near the target --- is present, and on the value of the coefficient of dynamic friction. We present a theory for the scaling of $A-B$ with the coefficient of friction that suggests that dissipation is primarily a perturbative effect in this flow, rather than the source of qualitatively different behavior.

\end{abstract}

\maketitle

\begin{figure}[t]
\includegraphics{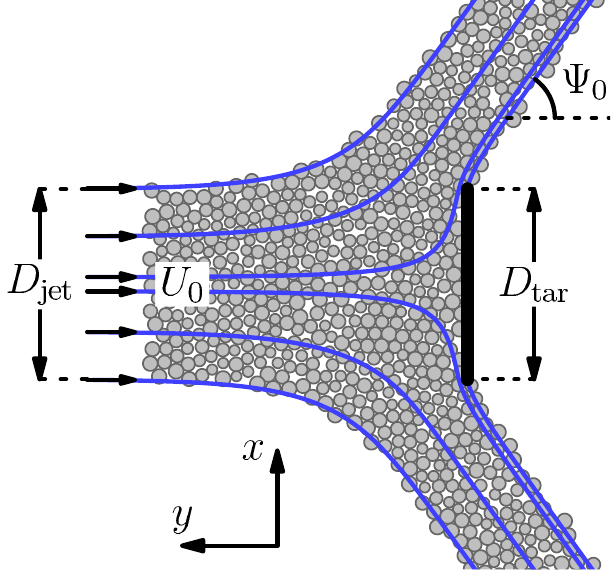}
\caption{Geometry of the granular jet impact. Grains are projected from the left at a target on the right, and form a cone with interior angle $\Psi_0$. In the granular experiments, gravity acted in the downwards direction but was negligible compared to the velocity scale of the impact. We do not simulate the effect of gravity in our simulations.}
\label{Fig1}
\end{figure}

In surprising experiments by Cheng, et al\cite{cheng2007collective}, it has been shown that a dense cohesionless granular jet impact can behave like a water flow of the same geometry. The experiment consists of a jet --- either granular or liquid --- being projected at some speed at a cylindrical target. The fluid forms a cone upon leaving the target, and the interior angle of the cone is measured (Fig.~\ref{Fig1}). The granular flow, despite being cohesionless, is observed to form a collimated cone when leaving the target. The interior angle of this cone quantitatively agrees with the cone angles observed when a water jet impinges upon a target of the same geometry\cite{clanet2001dynamics}. Furthermore, the Reynolds number of the water jet is very large, suggesting that dissipation is not involved in determining the angle of the cone in the water flow. The comparison between the water flow and granular flow is therefore even more surprising, as the granular flow has strong dissipation from inelasticity and friction. This suggests somehow that despite the presence of large dissipation in the granular flow, that dissipation is not strongly influencing the bulk properties of the flow.

This is in contrast to the majority of granular systems, in which the dissipation is critical in determining their behaviors. For example, granular shear flows exhibit ordering depending on their inelasticity\cite{alam2003first, alam2003rheology}, and continuum theory of granular flows has a singularity with respect to the dissipation\cite{jenkins1983theory}. Even more surprising, flow past an obstacle - virtually the same geometry as the jet impact experiment - has very different phenomena when the granular flow is more dilute, that seem more in line with the dissipation-dominated picture of granular flows. Granular flows impinging upon obstacles have been observed to form shockwaves both in experiment\cite{amarouchene2006speed} and simulation\cite{wassgren2003dilute}. The behavior of these shocks and the drag on immersed objects is seen to be dependent on inelasticity, whereas the cone angle observed in the dense granular flow seems to be completely independent of the inelasticity of the material used in the experiments\cite{cheng2007collective}. 

There is an additional aspect of the dense granular flow that makes its correspondence to water flows surprising: the granular jet forms a deadzone --- a region in which particles are trapped in a static arrangement --- near the target. This type of flow structure is seen in granular column collapse, in which a certain part of the column does not displace throughout its dynamics\cite{lajeunesse2004spreading, balmforth2005granular, staron2005study}. This is a manifestly granular behavior that does not appear anywhere in the corresponding liquid jet case. Changing the target shape influences the cone angle for water\cite{birkhoff1957jets}, but adding this dead region in the granular flow does not seem to alter the agreement between the water flow cone angle (which lacks such an internal shape), and the granular flow cone angle. One might expect that the deadzone would have a similar impact on the granular flow as a wedge would on the water flow, but instead it seems to have a much smaller effect. This paper attempts to address the effect of granular properties --- the deadzone and grain-grain dissipation --- upon the cone angle of the leaving flow, in order to understand why the granular and liquid flows are so broadly similar even though they are microscopically distinct.

Our tool for addressing this problem will be simulation. Simulations of the 2D granular jet impact have been performed using a molecular dynamics approach\cite{huang2010granular}. The grains in these simulations are of finite rigidity, which introduces a fast time scale to the problem (relative to the time scale due to relative motion between grains). In such simulations, this fast time scale limits the timestep used, and therefore the overall computational cost of the problem. We instead use a hybrid timestep driven rigid body collision method\cite{guttenberg2011approximate} that approximates perfectly rigid grains. The perfect rigidity is an approximation that we consider supported by the large difference in pressure scale in the experimental jet (on the order of 100 kPa, based on glass beads at 10 m/s\cite{cheng2007collective}) and the Young's modulus of the grains (on the order of 50 GPa for glass). 

In these simulations we can precisely vary the grain properties and the boundary condition in ways that are difficult to access experimentally. We expect that the cone angles observed in 2D simulations will not quantitatively agree with those observed in 3D experiments. However, we are interested in what various granular properties do to the cone angle, and we do expect such effects to be qualitatively the same in 2D and 3D, as we reproduce experimentally measured $A-B$ values with the 2D simulations. In order to compare to a corresponding fluid flow, we can use exact 2D solutions of the Euler equations\cite{zhang2011perfect}

\begin{equation}
\frac{d\vec{u}}{dt} + \vec{u} \cdot \vec{\nabla} \vec{u} = \vec{\nabla P}, \vec{\nabla} \cdot \vec{u} = 0
\end{equation}


First, we will address the general structure of the flow, including the deadzone. Granular material dropped onto a surface forms a static pile with a particular characteristic angle, the angle of repose, which is determined by the properties of the individual grains --- their shape and the grain-grain friction. We propose that the deadzone in granular jet impacts may be a similar feature. Rather than being held in place by gravity though, it is held in place by the pressure from the incoming granular jet. If this is the case, then a necessary element for the formation of the deadzone should be whether or not there is friction between the grains and the target. If the target is frictionless, force chains must come in perpendicular to its surface, and a static pile cannot be supported.

Next, we will address the effects of dissipation in the system. There are two sources of dissipation: the coefficient of restitution (inelasticity), and microscopic friction. 
The restitution coefficient appears to have no effect on the cone angle in experiments, and the friction is experimentally difficult to control. We use simulations to precisely control these parameters and investigate the behavior of the cone angle, and find that there are in fact small effects whereby the friction in particular can influence the cone angle. We observe inelasticity to have almost no effect on the cone angle unless there is absolutely no friction, and so we do not focus on it too heavily here. We propose that friction is playing a mostly perturbative role in this impact geometry, in which case we can predict the effect of changing the grain friction by computing the total amount of dissipation we expect in a homogeneous assemblage of individual grain-grain collisions. This then acts as an envelope for the amount of dissipation experienced by the granular jet, and tells us how the cone angle should change. We compare this theoretical prediction to measurements from our simulations and find good quantitative agreement.

\section{Cone Angle}


\begin{figure}[t]
\includegraphics[width=\columnwidth]{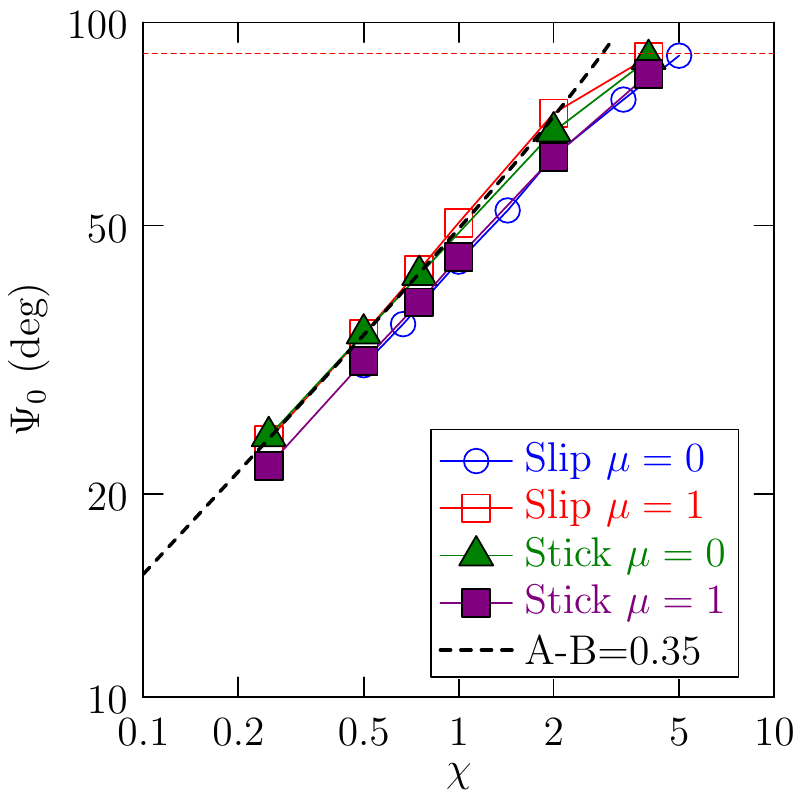}
\caption{Cone angle measured as a function of $D_t/D_j$ in 2D granular jet impact simulations. Unlike in 3D, the cone angle scales as the square root of the ratio of the target to the jet diameter. We show data for both stick and slip boundaries, at $\mu=0$ and $\mu=1$. The dashed black line is the theoretical curve for $A-B=0.35$. Note that despite the fact that the theoretical curve is for the asymptotic limit where $\chi \ll 1$, in practice the data follow this curve up until $\chi \approx 2$.}
\label{Fig2}
\end{figure}

The cone angle may be understood as the combination of two factors: transfer of momentum from the target to the jet, and dissipation that occurs between the inflow and outflow. A derivation for the functional form of the cone angle $\Psi_0$ in three dimensions appears in\cite{clanet2001dynamics}. However, care must be taken as the result is different in two dimensions. We present a version of the derivation here that takes into account an arbitrary dimension $d$. 

We begin with the flux of momentum parallel to the jet (the $y$ direction) per unit time. This is $q_y=Q U_0$, where $Q$ is the mass flux $Q\equiv \rho U_0 \sigma_j$ and $\sigma_j$ is the cross-sectional area of the jet and $U_0$ is the inflow speed. The target exerts a vertical force on the jet equal to $F_t = A \rho U_0^2 \sigma_t$, where $\sigma_t$ is the area of the target and $A$ is a dimensionless geometric factor. As such, the $y$ component of the leaving velocity is $U_y = (q_y - F_t)/Q$.

We now use conservation of energy to compute the leaving velocity. Some dissipation will occur associated with the deformation of the flow around the target. We assume that this dissipation is, to first order, proportional to the area of the target, and so $U_f \approx U_0 (1-B \frac{\sigma_t}{\sigma_j})$. This restricts us to situations in which the target area is much smaller than the scale of the overall flow. 

Putting these results together, we have an expression for $\Psi_0$:

\begin{equation}
\cos(\Psi_0) = \frac{1-A \frac{\sigma_t}{\sigma_j}}{1-B\frac{\sigma_t}{\sigma_j}}
\end{equation}

We define $\chi \equiv \frac{\sigma_t}{\sigma_j} = (\frac{D_t}{D_j})^{d-1}$, and Taylor expand the denominator:

\begin{equation}
\Psi_0 \approx \cos^{-1}(1-(A-B)\chi)
\label{ConeAngleEq}
\end{equation}

The asymptotic behavior of this equation as $\chi \rightarrow 0$ is $\Psi_0 \propto (A-B)\sqrt{\chi}$ (where $\chi = (\frac{D_t}{D_j})^{d-1}$). In 3D, this means that the cone angle asymptotes to a linear dependence on the ratio of the target diameter to the jet diameter, with a slope equal to $A-B$. In 2D, however, the cone angle asymptotes to a square root dependence on the target-to-jet ratio, with prefactor $A-B$. The factor $A-B$ describes the combined effect of flow geometry due to the target ($A$) and dissipation ($B$) and in the limit of small target to jet ratio is independent of the size of the jet with respect to the target, and so provides a useful quantity for characterizing the behavior of the cone angle across both 2D and 3D. 


In simulations, we measure this cone angle for $\chi<2$ by accumulating all grains that have moved past the target at least $200$ grain radii. We divide the resultant grains into those that have $x$ coordinates greater than the center line coordinate ('right'), and those that have $x$ coordinates less than the center line coordinate ('left'). We accumulate the average $y$ component of the velocity weighted by grain mass $\bar{v_y}$, and the average $x$ component of the rate of travel away from the centerline $\bar{v_x}$ (so grains right of the centerline contribute $v_x$, and grains left of the centerline contribute $-v_x$). In our coordinates, $\bar{v_y}<0$. We use these velocities to determine the cone angle $\Psi_0 = \pi-\tan^{-1}(-\bar{v_y}/\bar{v_x})$. When $\chi\geq2$, this technique runs into problems as much of the outflow remains above the target for the span of our simulation domain. In these cases, we must relax the $200$ grain radius cutoff, and instead use a cutoff that grains should have travelled at least $200$ grain radii along the direction of the target's surface before we consider them. 

We demonstrate the predicted 2D square root scaling in Fig.~\ref{Fig2}. Note that while the scaling only necessarily applies in the asymptotic limit as $\chi \rightarrow 0$, it is in practice obeyed over almost the entire range of behavior until $\Psi_0 \rightarrow 90^\circ$. As such, we may use $\chi=0.5$, which gives us a good combination of resolution of flow structures (improved by a large target) and resolution of the outflow cone (best for angles near $45^\circ$), and still extract $A-B$ from this relation. 


\section{Deadzone}

\begin{figure}[t]
\includegraphics[width=\columnwidth]{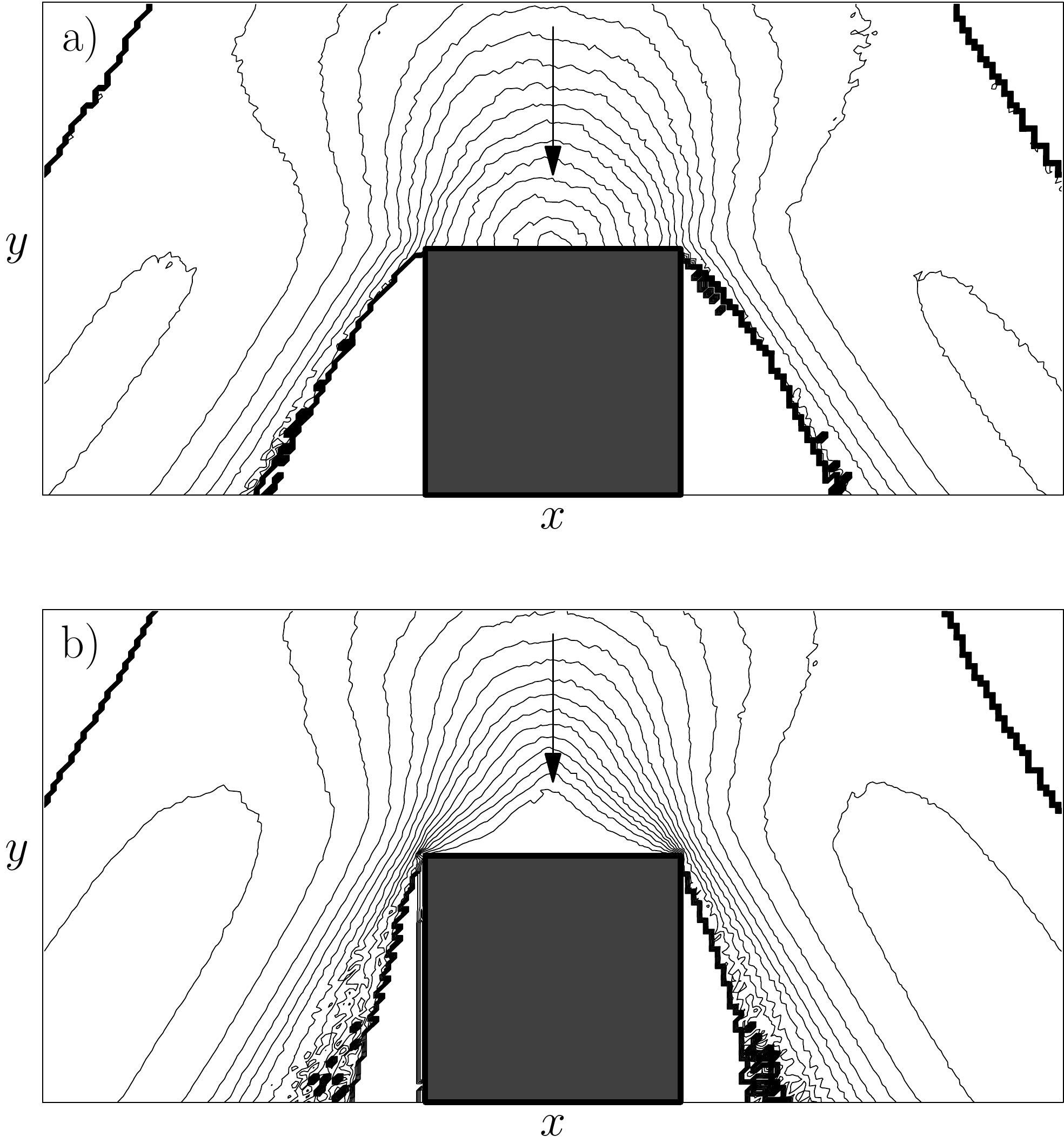}
\caption{Profiles of the L1 norm of the velocity for slip (a) and stick (b) boundary condition simulations with $\mu=1.0$ and $\chi=0.5$. In the simulation with a slip boundary condition, the velocity linearly approaches zero at the stagnation point, whereas in the simulation with the stick boundary condition there is a cusp-shaped deadzone within which the velocity decays exponentially. The flow is from the top, and the shaded square is the target. }
\label{Fig3}
\end{figure}

\begin{figure}[t]
\includegraphics[width=\columnwidth]{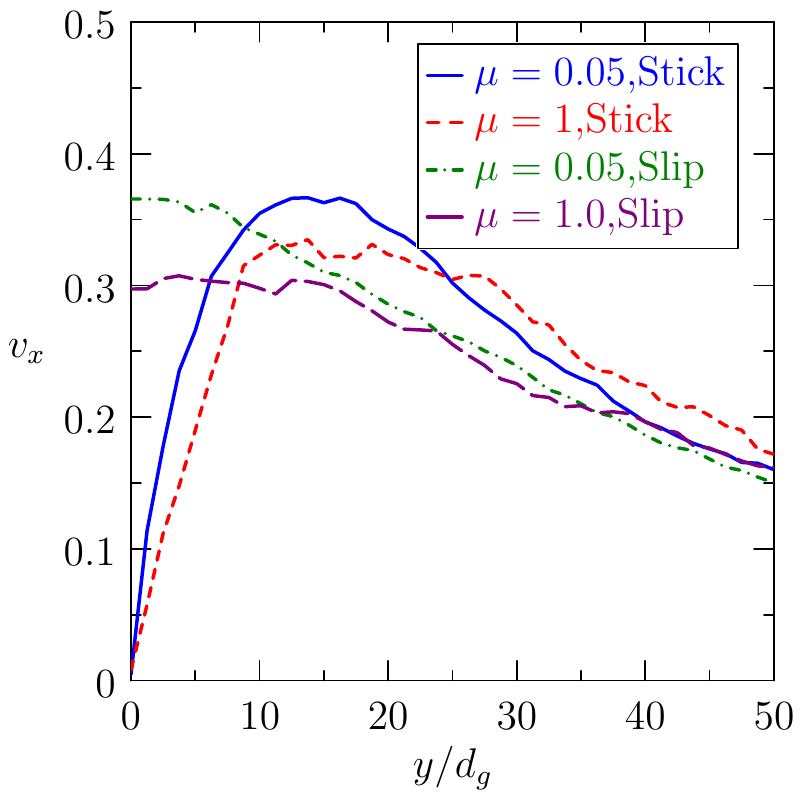}
\caption{Profile of the transverse ($x$) velocity along a line displaced $0.25 D_t$ from the centerline. Data are shown for two different values of the friction coefficient $\mu$, and for stick and slip boundary conditions. In the case of stick boundary conditions, the constraint that the velocity be zero at the target results in a local maximum in the transverse velocity.}
\label{Fig3b}
\end{figure}

\begin{figure}[t]
\includegraphics[width=\columnwidth]{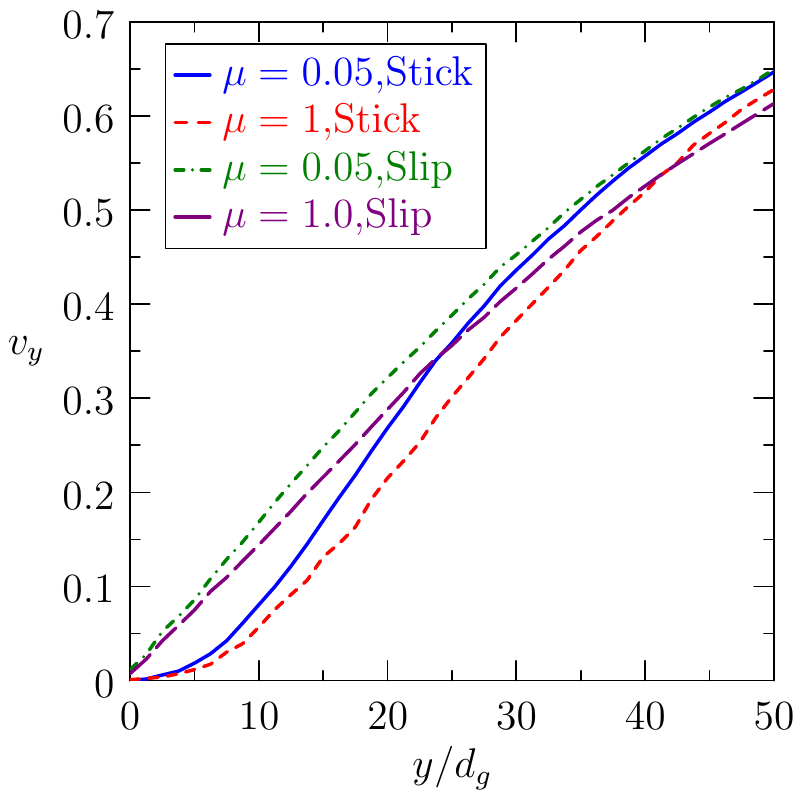}
\caption{Profile of the longitudinal ($y$) velocity along the centerline. Data are shown for two different values of the friction coefficient $\mu$, and for stick and slip boundary conditions. In the case of stick boundary conditions, the velocity decays exponentially towards zero near the target, whereas 
in the slip case it decays linearly. }
\label{Fig3c}
\end{figure}

Experimentally, a granular jet impinging upon a steel target produces a deadzone in which the particles are held in place by pressure from the incoming jet. In order to test whether the deadzone arises from tangential support at the target surface, we vary the boundary condition at the target in our simulations. We implement two different boundary conditions: 'slip' and 'stick'. In order to implement a slip boundary condition, we reverse the perpendicular velocity of grains that impact the target but otherwise leave them unchanged. The stick boundary condition corresponds physically to a rough target, one that is decorated with a layer of grains. We cause any grain that impacts the target surface to adhere, and so subsequent grains collide with the decorated target, using the collision rules for grain on grain collisions. These boundary conditions are the two extreme limits of a continuum of possibilities. In practice, the experimental target has some coefficient of friction for grain-target collisions, which may be different than that for grain-grain collisions. In order to more directly address the matter of the presence or absence of the deadzone, we ignore the extra degree of freedom here.

Our simulations consist of a jet one hundred grains in diameter, with grains being constantly injected into the simulation domain to form the jet, and deleted as they leave the simulation domain. When the system is fully developed, it contains approximately 140000 grains, varying somewhat with cone angle. As this is a system comprised of perfect hard spheres, the only dimensionless groups are $\frac{D_t}{D_j}$ and $\frac{D_t}{d_g}$ (where $d_g$ is the diameter of a single grain), and so our results are independent of the jet velocity and the grain density. We use a polydisperse set of grains uniformly distributed in radius between $0.8$ and $1.2$ to prevent crystallization. The grain density is held constant, and so the grain mass scales as the square of the grain radius. We initialize the jet by placing grains randomly in space in an inflow region and then allowing them to relax their positions to avoid overlap. The result is a packing fraction of $\phi=0.82$, close to jamming. In this paper we will generally discuss results for a fixed geometry $\frac{D_t}{D_j} = 0.5$. This is well within the scaling regime in which $\Psi_0$ obeys the theoretical prediction of Eq.~\ref{ConeAngleEq}. As shown in Fig.~\ref{Fig2}, we have done simulations at several target-jet ratios to determine the quality of this scaling assumption, and the data fall precisely upon the predicted curve at these values of $\chi$.


For our grain material properties, we choose a coefficient of restitution of $0.9$, which is consistent with the glass beads used in the experimental jet impact. However, we also find that our results do not significantly depend on our choice of restitution coefficient, as the frequency of collisions is large enough that almost all available energy is dissipated very quickly. We have Coulomb friction between the grains, with a coefficient of friction $\mu$ that we vary. Due to the presence of tangential forces, we track both the velocity and angular velocity of grains. For the purposes of observing the deadzone, we note that whether we obtain a deadzone or not is insensitive to our choice of $\mu$, which we have varied from $10^{-3}$ to $10$. This can be seen in Fig.~\ref{Fig3b} and Fig.~\ref{Fig3c}, which show that the general character of the velocity profiles is strongly dependent on the boundary condition, but only weakly depends on $\mu$.

We measure the time-averaged velocity profiles near the target for both stick and slip boundary conditions. We find that in the case of a stick boundary condition, the near-target velocity contours have a cusped structure. Within that contour, the velocity decays exponentially towards zero, with a decay length scale of a few grain diameters (Fig.~\ref{Fig3}b, Fig.~\ref{Fig3c}). This is consistent with the experimentally observed deadzone. However, when we switch to a slip boundary condition we observe no deadzone regardless of the grain-grain friction coefficient we choose. Instead we observe a stagnation point, where the velocity linearly approaches zero (Fig.~\ref{Fig3}a, Fig.~\ref{Fig3c}). Surprisingly, the velocity contours around this stagnation point quantitatively agree with the exact solution of the 2D Euler equation for this geometry, suggesting that even despite the high amount of dissipation in the system, the flow near the target is dominated by the interaction of geometry and the conservation of momentum\cite{zhang2011perfect}. 

\section{Friction Effects}

\begin{figure}[t]
\includegraphics[width=\columnwidth]{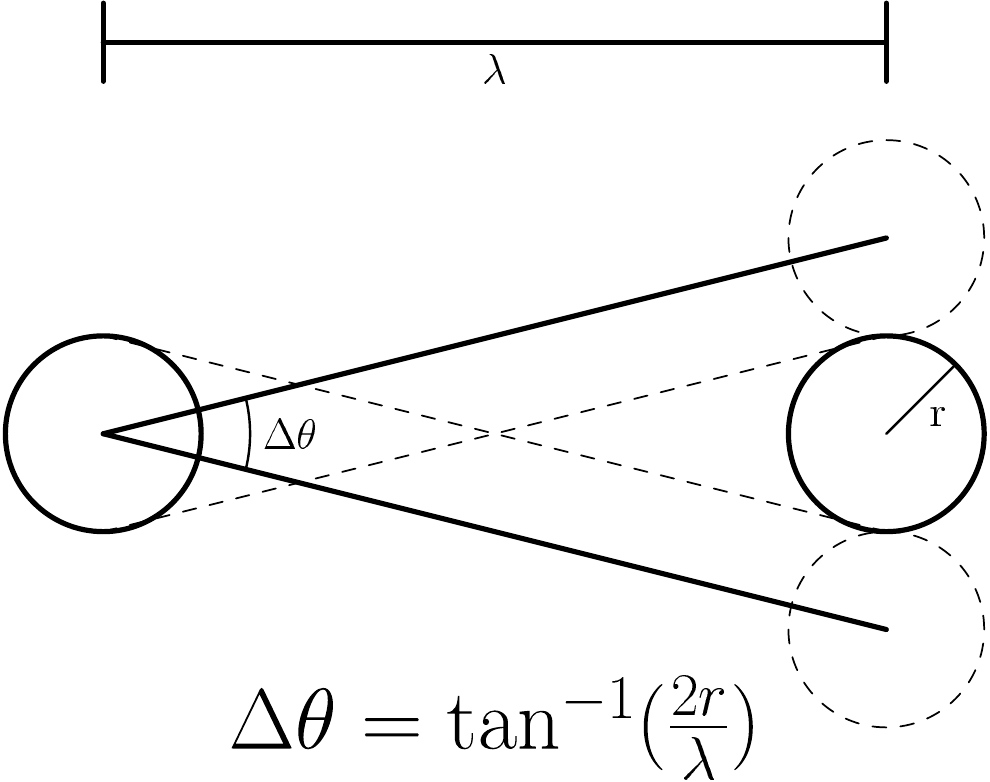}
\caption{Schematic of the set of angles of approach that will bring two grains starting at a distance $\lambda$ to collision.}
\label{Fig4}
\end{figure}


We now turn to the effect of dissipation. We observe almost no dependence of the cone angle on inelasticity in the system when there is also friction (Fig.~\ref{Fig7}b). However, the friction appears to have a weak but measurable effect. It is this effect that we will try to explain. In principle, friction could give rise to a change in deadzone geometry that alters the geometric contribution to the reaction force of the target on the jet. However, as we have shown in the previous section, this effect is very small (Figs.~\ref{Fig3b}, \ref{Fig3c}). As such, we will proceed by neglecting the influence of friction with deadzone geometry, and observe whether this approximation successfully explains the behavior of $A-B$.

Consider a gas of grains with random velocities. If we take two such grains, then the angle between the vector between them and their relative velocity will be random. We can then restrict ourselves to pairs of grains that will collide given their current trajectory. The result is that only a narrow band of angles will lead to a collision, but that within that band all angles should be equally likely. The further apart the grains, the narrower the band of angles. This situation is pictured in Fig.~\ref{Fig4}. The width of the angle distribution is a function of the mean distance between grains $\lambda$:

\begin{equation}
\Delta \theta = \tan^{-1}(\frac{2 r}{\lambda})
\end{equation}

We define the collision parameter $b$ to be the offset between the centers of the grain perpendicular to their relative velocity at contact, normalized by the sum of the grain radii. We can then express the fraction of energy lost in a collision as a function of $b$. We compute the distribution of collision parameters in random collisions in this granular gas, given an initial distance and uniform distribution over colliding values of $\theta$.

If the two grains approach at an angle $\theta$, then $b$ is:

\begin{equation}
b=\xi \sqrt{1-\eta^2} + \sqrt{1-\xi^2} \eta
\end{equation}

where $\xi \equiv \sin(\theta)$ and 

\begin{equation}
\eta \equiv \xi \left( \frac{\lambda}{2 r} \sqrt{1-\xi^2}-\sqrt{1-\left(\frac{\lambda}{2r}\right)^2 \xi^2} \right)
\end{equation}

Transforming the uniform distribution in $\theta$ to $P(b)$ gives us:

\begin{equation}
P(b) = \frac{1}{\Delta \theta} \frac{1}{|P^\prime (\theta(b))|}
\label{BDistribution}
\end{equation}

\begin{figure}[t]
\includegraphics[width=\columnwidth]{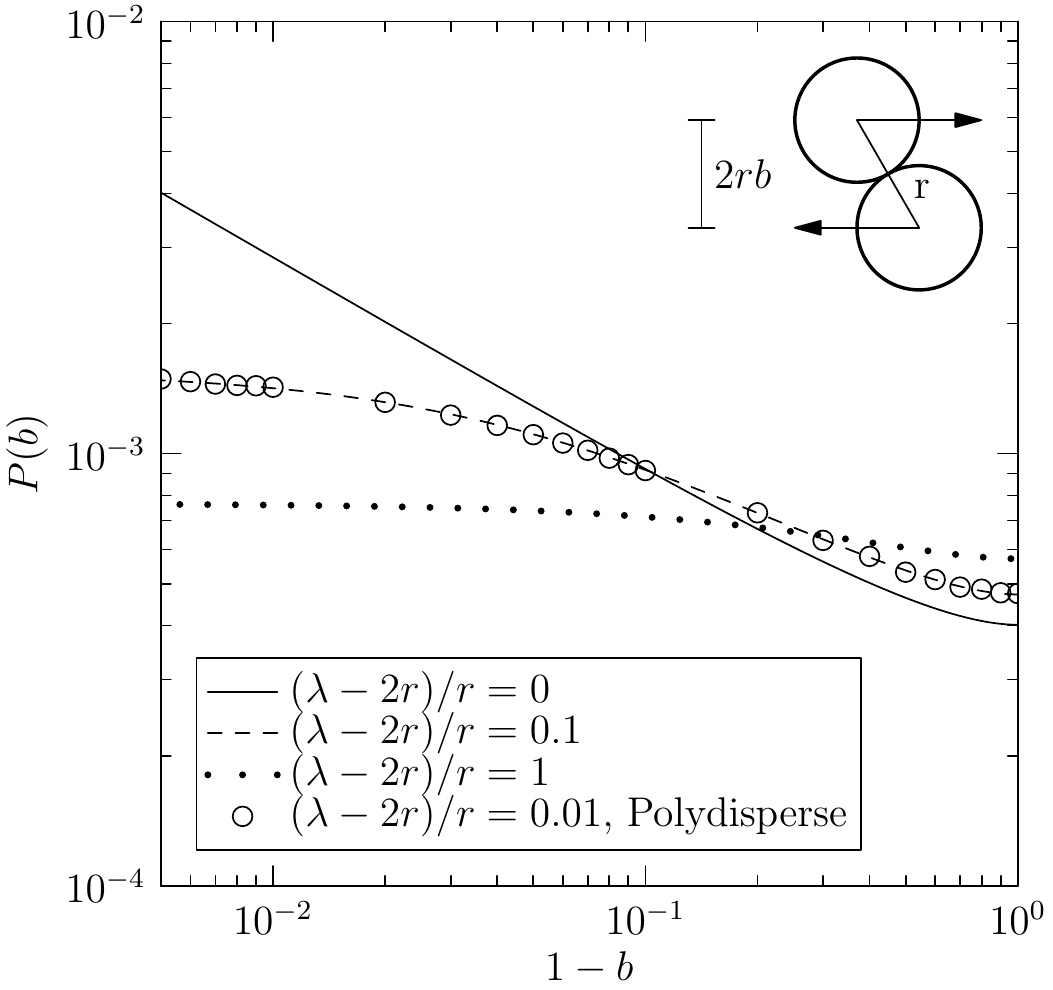}
\caption{Collision distribution of in terms of the collision parameter $b$ for different average distances between grains. When the average distance between grains approaches the sum of their radii, the collision distribution diverges with a square root singularity near $b=1$. Data from a mean-field collision simulation using uniformly distributed polydisperse grains of radii $[0.8,1.2]$ is shown for $\lambda = 2.1r$, and is indistinguishable from the mono-disperse distribution. }
\label{Fig5}
\end{figure}

We show $P(b)$ for various values of $\lambda$ in Fig.~\ref{Fig5}.

If we take the limit where $\lambda \rightarrow 2r$, then we have a situation in which grains are constantly in collision with eachother. This corresponds to a dense granular pack. While this limit is somewhat questionable, as now there can be a strong ordering in the pack, if the way that the pack is driven by exterior forces is random, the basic assumption of no correlation between the angle of the velocity of a grain and the angle of their contact still holds. In this limit, we observe that $b\rightarrow \sin(\theta)$, and so $P(b)$ asymptotically behaves as:

\begin{equation}
P(b)\propto\frac{1}{\sqrt{1-b^2}}
\end{equation}

This result also applies to polydisperse grains in contact, as the grains do not move between collisions and so the entirety of the distribution is determined by the random choice of relative velocity, and not the grain geometry. This can be seen to hold even away from the close-packed limit in Fig.~\ref{Fig5}, where we present data both from analytic calculation of the distribution and from a random sampling of $5 \times 10^7$ polydisperse collisions. For polydisperse grains, a constant $\lambda$ is not well defined, and so instead we hold constant the distance between the surfaces of the grains (corresponding to $\lambda - 2r$ in the monodisperse case). 


We now look at the energy dissipated in an individual collision occurring at collision parameter $b$, with coefficient of restitution $\epsilon$ and friction coefficient $\mu$. We assume that these grains have no relative rotation when they come into contact with eachother in order to simplify the calculation. 

The impulse felt by colliding grains can be separated into a normal component $I_N$ in the direction $\hat{r}$ and a tangential component $I_T$ in the direction $\hat{t}$. If two grains approach in the center of mass frame with a momentum $\vec{p_{i}}$ and no relative rotation, these components are:

\begin{equation}
|I_N| = (1+\epsilon) |\vec{p_{i}} \cdot \hat{r}|
\end{equation}

\begin{equation}
|I_T| = \min( \mu (1+\epsilon) |\vec{p_{i}} \cdot \hat{r}|, \frac{\vec{p_i} \cdot \hat{t}}{3} )
\end{equation}

The factor $\frac{1}{3}$ in $|I_T|$ assumes that the grains are homogeneous discs. In general, this factor is $\frac{1}{1+m r^2 / I}$, where $I$ is the moment of inertia of the grain. Observing that $\vec{p_{i}} \cdot \hat{r} = |p_{i}| \sqrt{1-b^2}$ and $\vec{p{i}} \cdot \hat{t} = |p_{i}| b$, we can express these impulses in terms of the collision parameter $b$. Given these impulses, we can compute the fraction of energy lost in the collision $\gamma \equiv 1-\frac{E_f}{E_0}$. 

If $b/3 < \mu (1+\epsilon)\sqrt{1-b^2}$, then:

\begin{equation}
\gamma_<(b) = (1-\epsilon^2)(1-b^2)+\frac{5b^2}{9}
\end{equation}

otherwise,

\begin{equation}
\gamma_>(b) = 2(1+\epsilon)(1-b^2+\mu b\sqrt{1-b^2})-(1+\epsilon)^2(1+\mu^2)(1-b^2) 
\end{equation}

This dissipation function is plotted in Fig.~\ref{Fig6}. As the coefficient of friction increases, only collisions closer and closer to $b=1$ are affected. However, for a dense pack the probability density of such collisions diverges, and so these glancing collisions are where the effect of higher friction dominates. We estimate the total amount of dissipation with the integral:

\begin{figure}[t]
\includegraphics[width=\columnwidth]{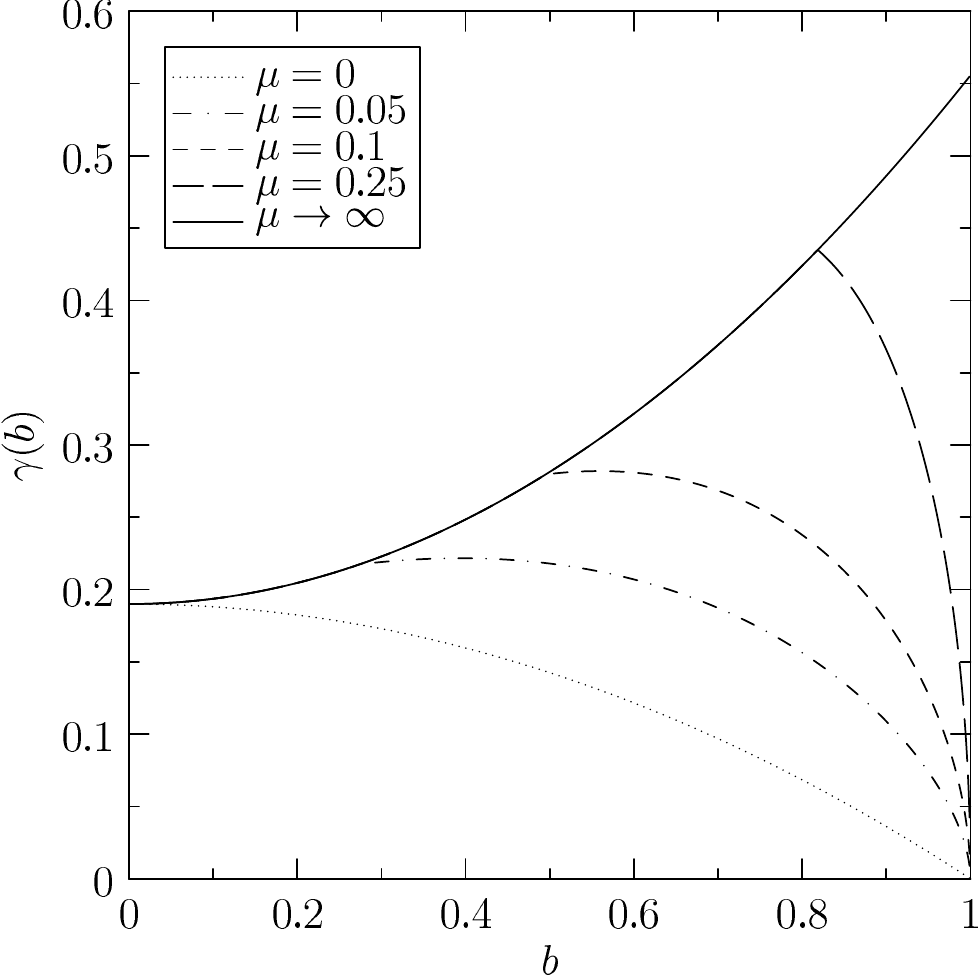}
\caption{This plot shows the fraction of energy dissipated in a collision as a function of the collision parameter $b$. }
\label{Fig6}
\end{figure}

\begin{equation}
D(\mu)=\int_{-1}^1 P(b) \gamma(b) db 
\end{equation}

We expect that this dissipation integral should control the dimensionless parameter $B$ in the quantity $A-B$. When $\mu\rightarrow 0$, we expect $A-B$ to be controlled mostly by the geometric contribution (although inelasticity still produces some finite dissipation). As such, this asymptotic value is not predicted by the present theory and is one fit parameter of this model. Secondly, the dissipation integral is an estimate of the fraction of energy dissipated in a set of uniformly distributed collisions, rather than a particular flow. As such, we expect the dissipation observed in the jet impact to scale with $D$ but not to be exactly equal to $D$. We therefore have a second fit parameter, which is a constant of proportionality that determines the influence of $D$ upon the quantity $A-B$. As such, we expect that:

\begin{equation}
A-B = C_0 - C_1 D(\mu)
\label{FitEq}
\end{equation}

To test this, we performed a series of simulations for different values of $\mu$, and measured $A-B$ in each case. We simulated both slip and stick boundary conditions to see if the presence of the deadzone made any difference in the scaling of $A-B$. The data are shown in Fig.~\ref{Fig7}a, along with fits of the dissipation theory to the data. The values of $\mu$ at which $A-B$ switches between its asymptotic behaviors are predicted by the theory with no adjustable parameters, as the two fitting constants do not influence the $\mu$ dependence of the curve, only the vertical scale. The first asymptotic regime seems to be fairly insensitive to the details of the collision distribution. However, we find that if we use a different $P(b)$, e.g. $P(b)\propto(1-b^2)^a$, then as $a\rightarrow 1$ the location of the second asymptotic regime goes to $\mu \rightarrow \infty$. 

We obtain agreement within errorbars for both the stick and slip boundary conditions. The effect of the target boundary condition is a fixed offset between the curves --- only the fit parameter $C_0$ differs between the two cases.

\begin{figure}[t]
\includegraphics[width=\columnwidth]{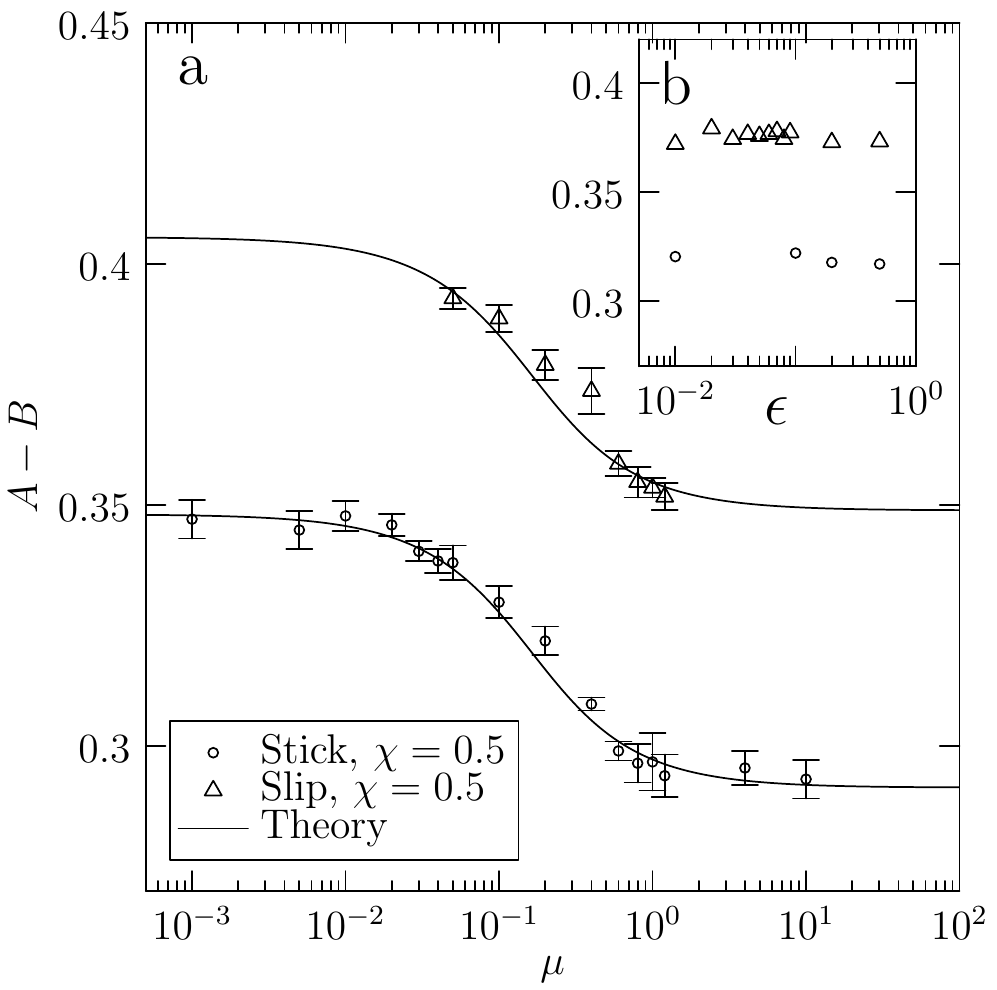}
\caption{a. This plot shows data for $A-B$ from simulations at $\chi=0.5$, along with fits to Eq.~\ref{FitEq}. The value of $A-B$ is determined by inverting the equation for the cone angle at a fixed value of $\chi$. The error bars are determined by measuring the cone angle at four different times during the simulation, and then computing the standard deviation of the resultant $A-B$ values. b. The inset shows the dependence of $A-B$ on the coefficient of restitution for a fixed value of $\mu=0.2$.}
\label{Fig7}
\end{figure}


\subsection{Polydispersity}

\begin{figure}[t]
\includegraphics[width=\columnwidth]{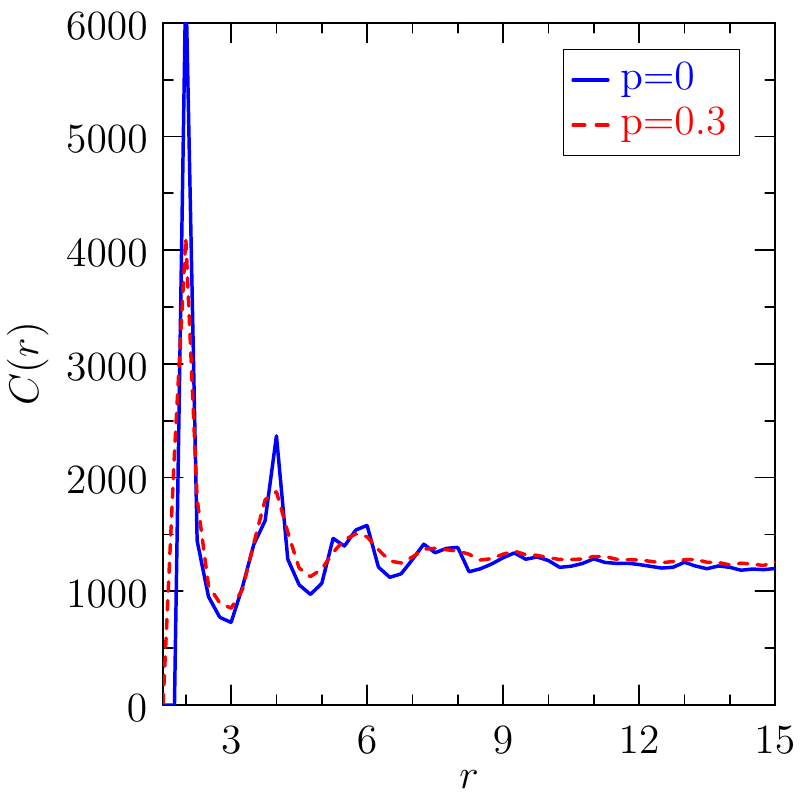}
\caption{This figure shows the correlation function of granular packs above the target for polydispersities $p=0$ and $p=0.4$. The effect of polydispersity is to decrease the correlation decay length.}
\label{Fig8}
\end{figure}

We also consider the matter of polydispersity. We generally use a polydisperse collection of grains to prevent crystallization (which can lead to markedly different behavior in granular systems). Our theory of the cone angle seems to be independent of any polydispersity in the system, though it is a mean-field theory of a random pack and would not be expected to capture anything depending on system-wide ordering. If polydispersity impacts the macroscopic behavior of the granular flow, it could be an interesting connection between mesoscopic order within granular systems (e.g. partial crystallization) and bulk behavior.

Our method for generating the initial jet is to insert grains at random in a volume of space well above the target. We continuously relax the pack during flight to remove the initially generated overlaps. As the jet has a free surface, all overlaps are eventually resolved by this process (there is no possibility of generating a pack that cannot somehow be relaxed). Our grains are generated with a uniform distribution of width $p$, which parameterizes the polydispersity. Normally we use $p=0.4$. 

In order to determine whether this polydispersity has an influence on $A-B$, we performed simulations with a stick boundary condition and $\mu=1$, but with polydispersity values $0$, $0.1$, $0.2$, and $0.3$.  We examine the internal structure of these packs above the target (at least $400$ grain radii up) to identify how the polydispersity is influencing the microscopic order. We compute a two-point correlation function $C(r)=\frac{1}{r}\sum_{ij} \delta(r_{ij}-r)$, treating each grain as a delta function at its center. The result is the correlation function of a solid with disorder (Fig.~\ref{Fig8}), with several peaks whose amplitudes decay exponentially. Measuring the decay constant of the exponential for each polydispersity we can extract the length scale of the ordering. We find that this length scale is at most $3.5$ grain radii even for the monodisperse pack, and decreases to $2.8$ for $p=0.3$. This suggests that ordering of the initial jet should not play a large role in the dynamics of the impact as the length scale does not depend strongly on the polydispersity. As we increase the polydispersity, we decrease the range of the microscopic order in our jet. However, even for monodisperse grains we never achieve a fully crystalline jet above the target due to our method of preparation.

\begin{figure}[t]
\includegraphics[width=\columnwidth]{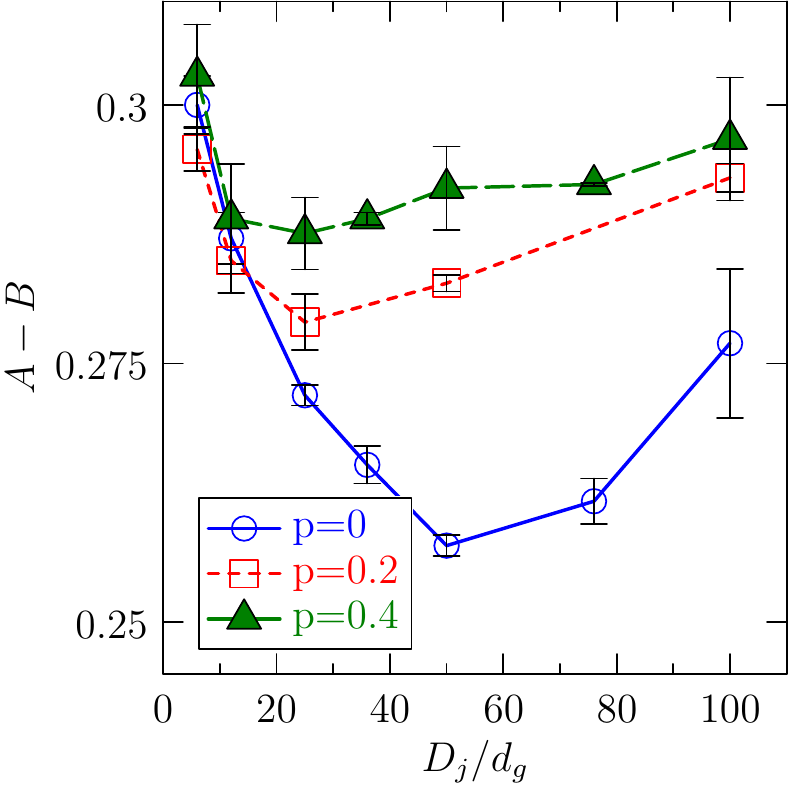}
\caption{This figure shows the effects of system size and polydispersity for a stick boundary flow at $\mu=1$. }
\label{Fig9}
\end{figure}

Because the local crystalline clusters have a different length scale than individual grains, we might expect that there could be a system size effect associated with the polydispersity. We perform simulations for various system sizes for polydispersities of $p=0$, $p=0.2$, and $p=0.4$ (Fig.~\ref{Fig9}). We find foremost that the monodisperse case is significantly different than the polydisperse cases, with a strong non-monotonic system size effect with a length scale of $D_j/d_g=50$. In the monodisperse case, the total range of the system size effect on $A-B$ is approximately $0.045$, which is comparable in scale to the effect of the presence or absence of a deadzone. Furthermore, it is not clear that the monodisperse case has reached its asymptote by $D_j/d_g=100$, and so further evolution of the behavior with respect to system size may occur.

For the polydisperse cases, $A-B$ increases with polydispersity. Aside from the monodisperse case, however, $A-B$ is not strongly influenced by polydispersity in the large system limit. The scale of the effect we observe is approximately $0.004$, an order of magnitude smaller than the friction and deadzone effects. Even in the polydisperse case, a finite size effect is observed, but with a length scale smaller than the monodisperse case (peaking at about $D_j/d_g=25$). The overall impact of system size on $A-B$ is also reduced in the polydisperse case, and it appears that $D_j/d_g=100$ is approaching the asymptotic limit with respect to system size (in the sense that the change of $A-B$ between a system size of $D_j/d_g=50$ and $D_j/d_g=100$ is about $0.005$, and so is about a $2\%$ on $A-B$). In the case of all polydispersities, the $A-B$ dependence on system size seems to follow the same curve for very small systems. At these small values of $D_j/d_g$, the outflow is no longer coherent. This suggests that the behavior is dominated more by random scattering than by bulk flow. As the system size increases, the flow becomes collimated (around $D_j/d_g=25$), and the curves diverge from eachother. Our current model does not predict the polydispersity effect at all, but it seems clear that it is a higher-order phenomenon that may not be capturable from a simple mean-field approach. 

\section{Conclusions}

We have shown that the behavior of a granular jet impinging upon a target is well described by the combination of a boundary condition effect at the target to capture the geometrical effects of the target reaction force, and a microscopic model for the fraction of energy dissipated in collisions within the jet. The boundary condition determines whether or not a deadzone forms, which effectively changes the target geometry. A boundary condition capable of supporting horizontal stresses is needed in order to observe a deadzone, whereas a slip boundary condition does not produce a deadzone even if the friction between grains is large. 


The presence or absence of the deadzone is responsible for a fixed offset in the value of the constant $A-B$. On the other hand, friction between the grains seems to be well-described as a local, homogeneous effect, changing only the total dissipation within the jet and not strongly influencing the large-scale structure of the flow. While we see only a small direct change to the velocity profiles due to friction effects (Figs.~\ref{Fig3b} and \ref{Fig3c}), not corresponding to any macroscopically altered flow structures, we observe a change in cone angle consistent with a microscopically homogeneous frictional effect. We have presented a calculation that estimates the total dissipation within the jet as a function of the friction coefficient, and find that it predicts the crossover as a function of $\mu$ between the asymptotic regimes of $A-B$ corresponding to no friction and infinite friction. 

It is not a given that friction should behave in this way. In general, factors such as surface tension or viscosity are associated with length scales, which give rise to corresponding extended structures in the flow (e.g. boundary layers) along with any homogeneous effects they may have. In the theory for the water jet, the influence of these factors on the cone angle is restricted by going to the limit where the length scales associated with surface tension and viscosity are much smaller than other length scales in the system, and so their effects have asymptotically saturated\cite{clanet2001dynamics}.

In contrast, friction and inelasticity do not create any new flow structures and do not introduce any new length scales in this flow geometry. As such, the effects are well-captured by the homogeneous picture. If we consider other granular systems, this approach should work so long as those systems do not develop boundary layer-like structures as a consequence of friction. The success of this microscopic description suggests an explanation for the correspondence between perfect fluid flow and the granular dynamics reported in \cite{zhang2011perfect}. The effect of dissipation is a weak perturbation and does not qualitatively affect the flow behavior. 

\section{Acknowledgements}

We would like to acknowledge Wendy Zhang, Heinrich Jaeger, Sid Nagel, and Jake Ellowitz for helpful discussions and suggestions. This work was funded by a University of Chicago NSF MRSEC Kadanoff-Rice Fellowship.

\bibliographystyle{apsrev}

\bibliography{bibliography}

\end{document}